# Overcoming the quantum limit of optical amplification in monolithic waveguides

Zhichao Ye†, Ping Zhao†, Krishna Twayana, Magnus Karlsson, Victor Torres-Company, Peter A. Andrekson*

† These authors contributed equally to this work

Department of Microtechnology and Nanoscience, Chalmers University of Technology, 41296 Gothenburg, Sweden

*Corresponding author. Email: peter.andrekson@chalmers.se

**Abstract:** Optical amplifiers are essential in numerous photonic applications. Parametric amplifiers, relying on a nonlinear material to create amplification, are uniquely promising as they can amplify without generating excess noise. Here, we demonstrate amplification based on the 3$^{rd}$ order nonlinearity in a single chip, while in addition reporting a noise figure significantly below the conventional quantum limit when operated in phase-sensitive mode. Our results show the potential of nanophotonics for realizing continuous-wave parametric amplification that can enable applications in optical communications, signal processing and quantum optics across a wide range of frequencies.

Optical parametric amplifiers (OPA) rely on a nonlinear material to create amplification, in contrast to stimulated emission as in conventional amplifiers. Due to their unique characteristic of providing a noise figure (NF) well below the 3 dB quantum limit (*1*) of conventional amplifiers when operated in a phase-sensitive mode (*2*), OPAs have attracted much interest in optical communication (*3*), ultra-fast signal processing (*4*, *5*) and quantum metrology (*6*). To date, continuous wave (CW) operation has only been demonstrated in bulky systems, for example using hundreds of meters of $\chi^{(3)}$-based highly nonlinear fiber (*3*) or using multiple $\chi^{(2)}$-based lithium-niobate planar waveguides which require periodic poling to generate phase matching and lossy free-space assemblies in between (*7*). In the past decades, extensive research has explored materials with high nonlinearity such as silicon (*8-10*), AlGaAs (*11*), nonlinear glasses (*12*, *13*), graphene (*14*), and plasmonics (*15*). However, all demonstrations to date using a single compact waveguide for amplification have operated with a pulsed pump and not in a CW operation prohibiting use in real applications. The reason for this is either waveguides with prohibitively high losses (*8-11*) or the use of materials that create large nonlinear losses (*12-16)*, such as two-photon- or free-carrier-absorption that will limit the pump power that can be used. Here, we address this challenge, by employing a very low-loss (1.4 dB/m) silicon-nitride waveguide within a chip area of 25 mm$^2$ to demonstrate CW parametric amplification of 9.5 dB with a noise figure (NF) of 1.2 dB, well below the conventional 3 dB quantum limit.

Silicon nitride features a relatively large nonlinearity and does not suffer from nonlinear loss at telecommunications wavelengths (*17*). Its index contrast to the silica cladding simultaneously allows high confinement, low loss and high power-handling ability by advanced nano-fabrication techniques (*18–20*). Despite the success in high-Q microresonators (*21*), low-loss (< 1dB/m) meter-long Si$_3$N$_4$ waveguides have only been achieved with low confinement (*22*), which is not suitable for OPAs due to their weak nonlinearity. To achieve CW-pumped parametric amplification based on four-wave mixing (FWM) in $Si_3N_4$, dispersion-engineered long waveguides with both high nonlinearity and low propagation losses are needed. An estimation of this can be made by noting that the maximum parametric gain G ≈ exp{$2\gamma P$[1-exp($-\alpha L$)]/$\alpha$}·exp($-\alpha L$) (*23*), where $\gamma$ is the Kerr nonlinearity parameter of the waveguide (~ 1 $m^{-1}W^{-1}$ can be achieved in high confinement $Si_3N_4$ waveguides), *P* is the pump power and $\alpha$ is the linear attenuation coefficient (*24*). The waveguide loss is a key factor since it directly impacts both the gain and NF (*25*). A schematic illustration of a waveguide OPA based on degenerate four-wave mixing is depicted in Fig. 1a. The input signal wave is amplified, and an idler wave is generated along the propagation, while simultaneously, the pump wave is attenuated.

We fabricated high-confinement dispersion-engineered $Si_3N_4$ spiral waveguides using an advanced subtractive processing method (*19*). A photograph of one of the fabricated chips with nine $Si_3N_4$ waveguides is shown in Fig. 1b. We cascaded 25 spiral waveguide units (shown in Fig. 1c) to construct a 1.42-meter-long waveguide within a device area of 25 $mm^2$. Stitching error compensation was implemented to overcome the drawback of limited writing fields with electron beam lithography (EBL), see Supplementary for detailed information. Optical frequency domain reflectometry (OFDR) was used to characterize the fabricated $Si_3N_4$ waveguide with a cross-section dimension of 690 nm x 2000 nm, with the result shown in Fig. 1d. No significant defect along the 1.42 m $Si_3N_4$ waveguide was observed, which verifies the success of the stitching error compensation. The waveguide loss achieved for transverse electric (TE) polarization was 1.4 dB/m, which is the lowest loss in high-confinement meter-scale $Si_3N_4$ waveguides to date. Twelve out of 20 fabricated waveguides were defect free according to the OFDR measurements. The mean propagation loss of the defect-free waveguides was 1.7 dB/m. The dispersion of the fundamental TE ($TE_{00}$) mode was designed to be anomalous, thus being suitable for parametric amplification.

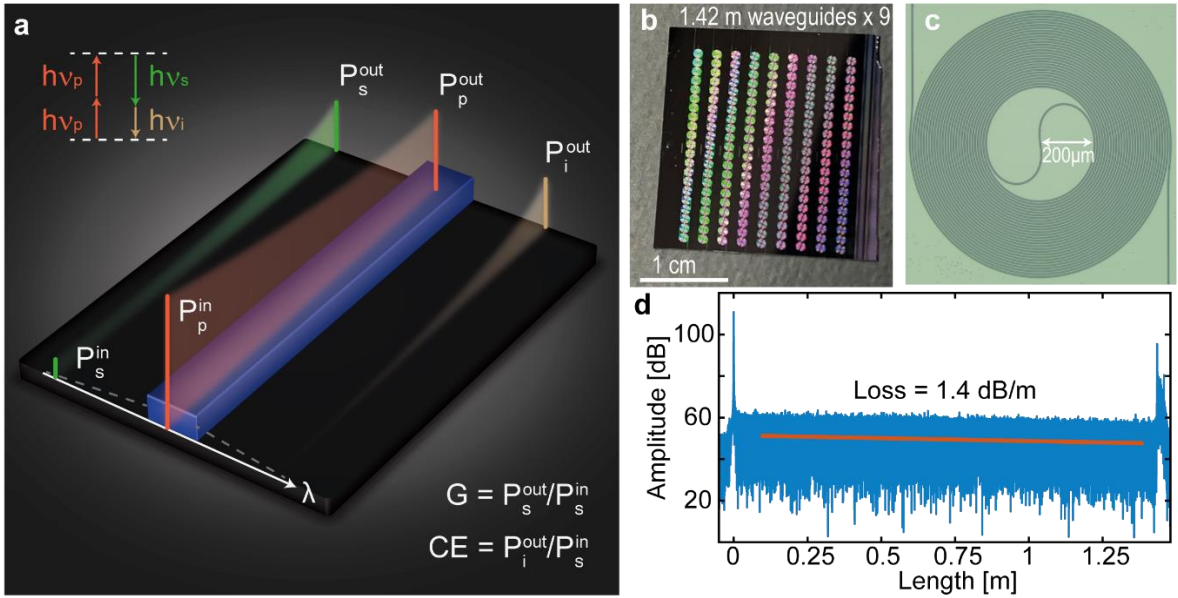

**Fig. 1. Schematic illustration of parametric amplification, $Si_3N_4$ chip and performance characteristics. a**, Schematic illustration of OPAs based on degenerate four-wave mixing. The green, orange and light brown lines represent signal, pump and idler waves, respectively. **b**, Photograph of a $Si_3N_4$ chip containing 9 waveguides, each with a length of 1.42 m. **c**, Optical

microscope image of one spiral waveguide unit. **d**, OFDR trace of one $Si_3N_4$ waveguide. The measured propagation loss of TE polarization is 1.4 dB/m.

We performed CW-pumped parametric amplification experiments using the $Si_3N_4$ waveguide in both phase-insensitive and phase-sensitive mode. Fig. 2a shows a schematic of the experimental setup (see Supplementary). For the phase-insensitive amplifier (PIA), only a signal and a CW 1563 nm pump were combined and injected into the waveguide. The PIA gain at the phase matched signal wavelength was simulated using the nonlinear Schrödinger equation (NLSE) using a 1.42 m long waveguide with a loss of 1.4 dB/m. Fig. 2b shows the resulting PIA gain which increases exponentially with the pump power, reaching 20 dB with a pump power of 34.4 dBm. Typical measured on-chip PIA gain and conversion efficiency (CE) spectra are shown in Fig. 2c. The corresponding pump and signal powers at the waveguide input were estimated to be 34.4 dBm and -22.5 dBm, respectively. Here, the on-chip PIA gain (CE) is defined as the ratio of the on-chip signal (idler) power at the output relative to that at the input. The obtained maximum on-chip PIA gain and CE were 6.4 dB and 5.3 dB, respectively. To the best of our knowledge, this is the first experimental demonstration of both on-chip PIA gain and net CE in an integrated $\chi^{(3)}$ waveguide with a CW pump. Moreover, the experimental PIA gain and the CE spectral profiles agreed with simulated ones based on NLSE with a pump power of 31.5 dBm and a fitted dispersion $\beta_2$ = -45 ps$^2$/km. More details on the simulations are presented in the Supplementary. Here, the effective $TE_{00}$ pump power is 2.9 dB smaller than the estimated actual on-chip input pump power. We believe that the difference is mainly due to the coupling to higher-order modes, and uncertainties in the nonlinear coefficient and coupling efficiency. The bandwidth over which the PIA provides gain is 28 nm and can be broadened or moved to another wavelength band with modified dispersion engineering (*26*). The inset in Fig. 2c shows the on-chip signal output power at 1553 nm versus on-chip input signal power ($P_s$). Excellent linear parametric amplification of the signal is observed. In order to test the ability of the $Si_3N_4$ waveguide used in optical communication, we carried out a wavelength conversion experiment of 10 Gbps non-return-to-zero signal. Due to high CE, no additional EDFA was needed to pre-amplify the idler before optical detection. The penalties for outcoupled signal and converted idler were 0.3 dB and 1 dB at a bit-error-rate of $1 \times 10^{-6}$, respectively (see Supplementary).

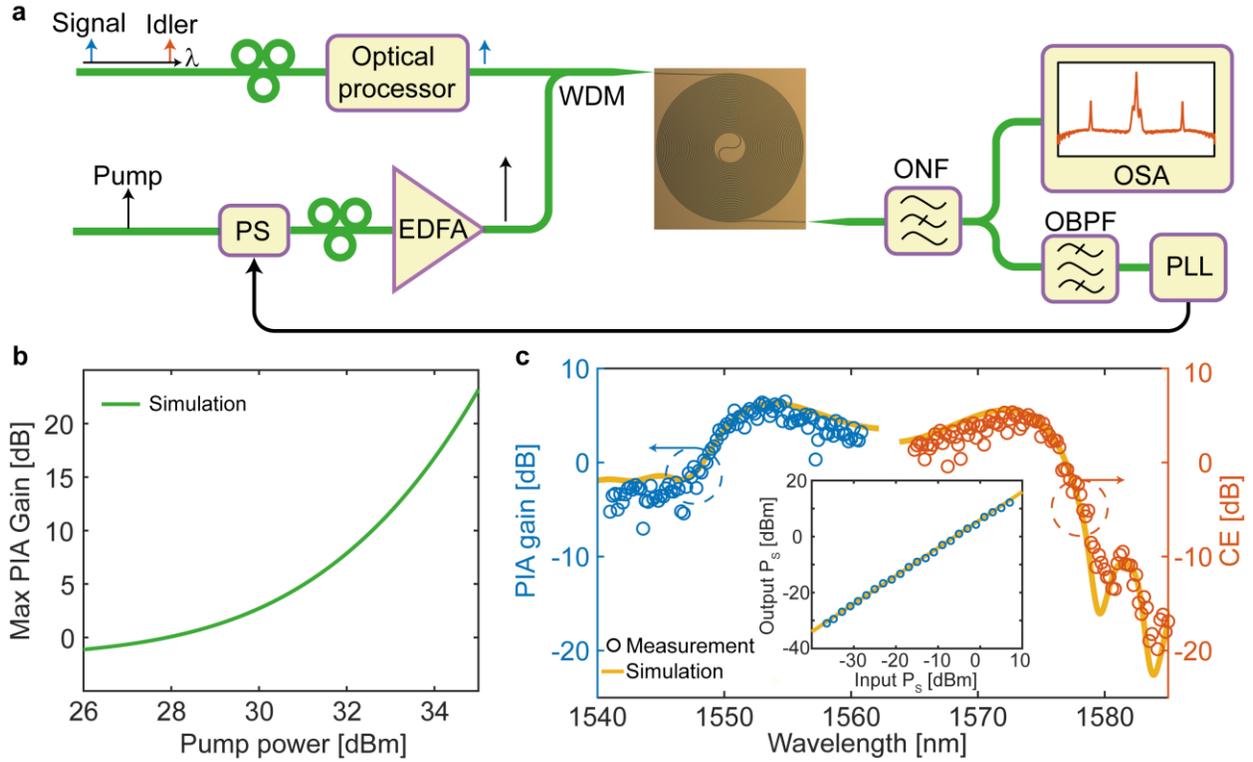

**Fig. 2. Parametric amplification. a**, The PIA and PSA experimental setup. WDM, wavelength division multiplexer; PS, phase shifter; OSA, optical spectrum analyzer; ONF, optical notch filter; OBPF, optical bandpass filter; PLL, phase lock loop. **b**, Simulated maximum PIA gain versus pump power. **c,** Spectra of on-chip PIA gain (blue circles) and conversion efficiency (red circles) with an on-chip pump power 34.4 dBm. The simulated PIA gain and CE (solid yellow lines) spectra are with a pump power of 31.5 dBm on $TE_{00}$ mode. The inset figure shows the measured output signal power versus input signal power with the linear fitted slope indicating 6 dB gain.

As a next step, we investigated phase-sensitive amplification (PSA) using the $Si_3N_4$ waveguide. We coupled 34 dBm pump power at 1563 nm into the chip, and the powers of the input signal (1554 nm) and idler (1572 nm) were -26.5 dBm each. The output spectra of the PSA and PIA are shown in Fig. 3a. As can be seen, the PSA provides 5.2 dB additional gain at 1554 nm compared to the PIA due to the coherent superposition of amplified signal and idler (*27*). Figure 3b shows the measured PSA gain at five signal wavelengths. The maximum PSA gain was 9.5 dB at 1554 nm which corresponds to nearly optimum phase matching. Additionally, we measured the NF of both PIA and PSA with the same pump power as presented in Fig. 3c. Also shown with the lines are the NF from theory for both PIA and PSA in the ideal case (no waveguide loss) and in the present case (1.4 dB/m waveguide loss). The PSA NF in the ideal case is 3 dB for 0 dB PSA gain, which is due to the co-injection of signal and idler, while only the signal is recovered at the output (*28*). Note that a theory for the NF of a PSA with a lossy waveguide has not been described earlier, and the details of this can be found in the Supplementary. The measured PIA NF at signal wavelength of 1556 nm is 3.3±0.4 dB. Theoretically, the PIA NF reaches a maximum of 3.6 dB at a gain of 7 dB and drops to about 3.4 dB at 25 dB PIA gain, as the amplified excess optical noise then dominates over the waveguide loss in terms of contribution to the overall NF. For the PSA, the measured on-chip NF is about 1.2±0.4 dB at 1556 nm. We note that the gain and NF reported above is the chip gain. When including the input and output coupling losses (2.5 dB/facet), the fiber-to-fiber gain and NF for the PSA were 4.5 dB and 3.7 dB, respectively. Figure 3d shows the normalized PSA gain with varying signal phase. The extinction ratio between maximum and minimum gain, which is an important factor for optical regeneration and squeezing (*4*), was 20 dB and, consequently, is promising in quantum optics (*29*).

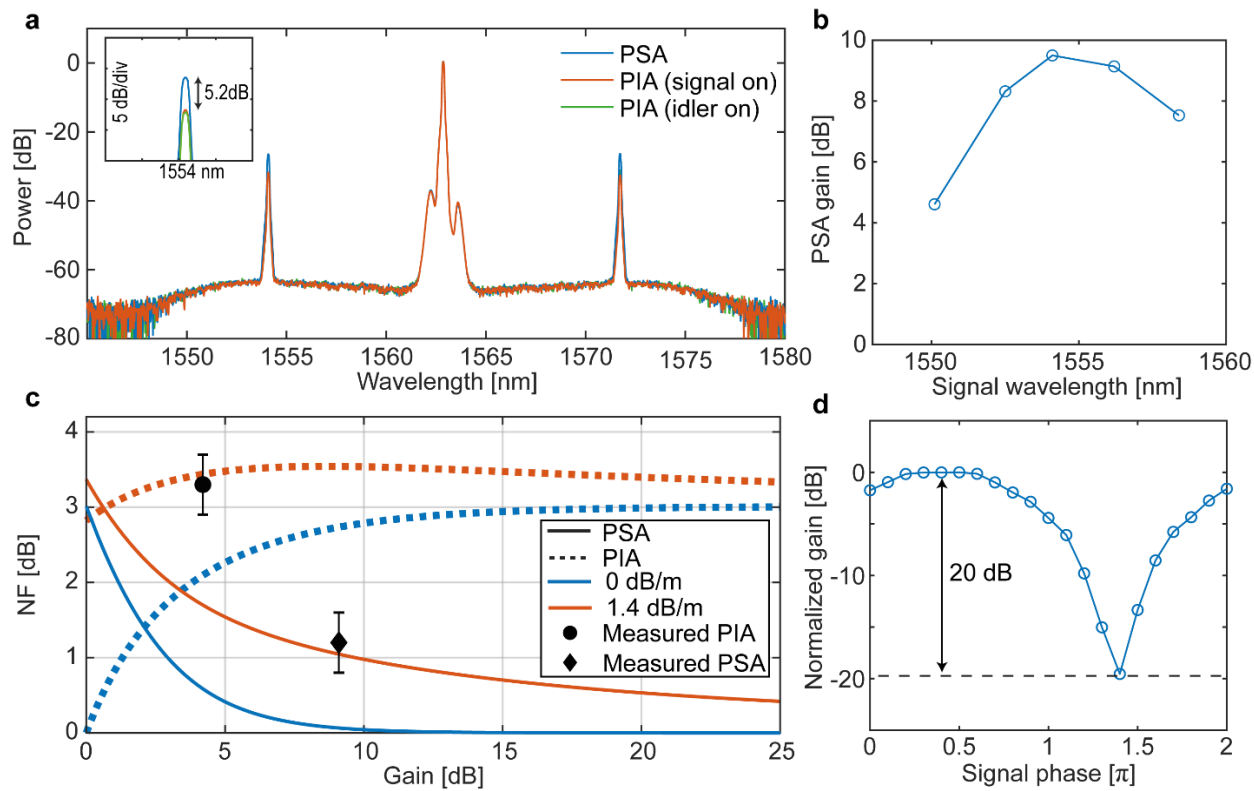

**Fig. 3. PSA characterization. a**, Output spectra of PSA (blue line), PIA with only signal on (orange line) and PIA with only idler on (green line). The inset figure shows the additional 5.2 dB gain observed with the PSA. **b**, On-chip PSA gain versus signal wavelength. **c**, Measured and theoretical results of on-chip NF for the PIA and the PSA. The solid blue and orange curves are the theoretical results of PSA with 0 dB/m and 1.4 dB/m loss, respectively, in the 1.42 m long waveguide. The dashed blue and orange curves are the theoretical result of PIA with 0 dB/m and 1.4 dB/m loss, respectively. **d**, Normalized phase-sensitive gain with varying signal phase.

In summary, we have for the first time experimentally demonstrated a CW-pumped as well as ultralow-noise OPA in an integrated photonic waveguide in the telecommunication band. The achieved gain and NF represent a milestone towards chip-scale optical signal processing and excess-noise-free amplification. We expect that lower NF with higher OPA gain can be achieved by further reducing the waveguide losses, increasing the waveguide length (*26*), and reducing the crosstalk between the fundamental and higher order optical modes. For practical applications, minimizing the coupling loss between the fiber and waveguide is also important as it leads to higher on-chip pump power and a corresponding exponential increase in OPA gain, as well as a reduced fiber-to-fiber NF. It should be noted that the OPA is scalable to other wavelengths since silicon nitride is transparent from the visible to the mid-IR wavelength range. With the advantages of low noise and a small monolithic footprint, CW-pumped silicon-nitride-based OPAs may open new possibilities in optical communications (*30*), ultrafast spectroscopy (*5*), quantum optics and metrology (*6*).

**Acknowledgments:** This work was funded by the Swedish Research Council (grant VR-2015-00535, VR-2020-00453), the K.A. Wallenberg Foundation and the H2020 Marie Skłodowska-Curie Innovative Training Network Microcomb (GA 812818). The authors thank Dr. Marcus Rommel for the fruitful discussion and assistance in EBL, Rasmus Larsson and Dr. Fuchuan Lei for their assistance in experiments. The $Si_3N_4$ samples were fabricated at Myfab Chalmers.

**Funding:**

Swedish Research Council VR-2015-00535 (PAA)

Swedish Research Council VR-2020-00453 (VTC)

K.A. Wallenberg Foundation, KAW Scholar (PAA)

H2020 Marie Skłodowska-Curie Innovative Training Network grant GA 812818 (VTC)

**Author contributions:**

Conceptualization: PAA, VTC, PZ, ZY

Methodology: PZ designed the OPA, figured out the waveguide dispersion and length for fabrication. ZY designed the waveguide layout and developed the nanofabrication process. PZ and ZY constructed the OPA setup. KT designed the OFDR for characterizing the waveguide loss

Investigation: PZ and MK performed the theoretical modeling and simulating of the OPA. ZY performed the fabrication of the waveguides. PZ and ZY conducted the OPA experiment. ZY, KT and PZ characterized the waveguides. PZ performed the wavelength conversion of modulated optical signal.

Data processing: ZY, PZ

Funding acquisition: PAA, VTC

Project administration: PAA

Supervision: PAA, VTC

Writing – original draft: ZY, PZ

Writing – review & editing: PAA, VTC, MK

**Competing interests:** Authors declare that they have no competing interests.

**Data and materials availability:** All data is available in the manuscript or the supplementary materials.

## Supplementary information

### Design, fabrication and characterization of spiral waveguides

The design of the waveguide follows an Archimedean spiral so that the curvature of the waveguide varies slowly and continuously along the propagation length. The S-bend connecting clockwise and anti-clockwise spirals, and the connection between adjacent spiral waveguide units are designed using an algorithm to minimize the variation of the curvature (*1*). This adiabatic bend design helps to reduce the coupling from $TE_{00}$ mode to higher order modes or radiation modes in the waveguide. The spiral waveguide unit was designed to fit within an 1x1 mm$^2$ EBL writing field (WF) in order to reduce the number of stitching errors present at the border of WFs. The WFs were arranged so that there was only one stitching error at the waveguide connecting the two adjacent spiral units. The spiral waveguide unit, the arrangement of WFs and the location of stitching error are depicted in Fig. S1a. Typically, the stitching error of 1x1 mm$^2$ EBL WFs is tens of nanometers, which can result in significant scattering loss. We noticed that our stitching error is systematic at every writing field border, indicating that the stitching error is caused by the imperfect calibration of the WFs (*2*). Thus, we actively compensated the stitching error by precisely shifting the adjacent WFs according to the regular EBL without any stitching error compensation. The SEM images of the exposed resist with and without stitching error compensation are depicted in Fig. S1b. The regular EBL exposure leads to significant stitching errors in both vertical and horizontal direction, while no obvious offset is observed for EBL exposure with stitching error compensation. $Si_3N_4$ waveguides were fabricated using the process detailed in (*3*). Multi-pass EBL was implemented to further reduce the sidewall roughness and reduce the waveguide loss. The $Si_3N_4$ waveguide was annealed at 1200°C under argon ambient atmosphere to reduce the absorption loss caused by N-H bonds. The wafers were manually cleaved into chips.

The OFDR measurements were carried out as in (*4*). The reflected light was monitored while the tunable laser was swept from 1520 nm to 1600 nm with 8 nm/s scanning rate, and the wavelength of the tunable laser was calibrated by a self-referenced fiber frequency comb (Menlo Systems FC-1500 with repetition rate of 250 MHz). The average coupling loss between the lensed fiber and the $Si_3N_4$ waveguide was calculated by subtracting the measured propagation loss (2.0±0.1 dB) based on OFDR from the total insertion loss (fiber to fiber, 7.0±0.3 dB). Thus, the coupling loss was estimated to be 2.5±0.2 dB/facet, considering equal coupling loss at both input and output facets.

### Simulation of PIA with a $Si_3N_4$ waveguide

The propagation of the optical field along a $Si_3N_4$ waveguide is governed by the NLSE as,

$$\frac{\partial E}{\partial z} + \frac{\alpha}{2} E + \frac{i\beta_2}{2} \frac{\partial^2 E}{\partial T^2} = i\gamma |E|^2 E \qquad (1)$$

where *E* is the slowly varying envelope of the overall optical field, $\beta_2$ is the group velocity dispersion coefficient at the pump wavelength, *z* is the optical propagation axis and *T* is the relative time axis where a reference frame moves with the envelope at the group velocity (*5*). We considered third-order optical nonlinear effects, as represented by the nonlinear coefficient. Raman and Brillouin scattering, higher-order dispersion, as well as nonlinear optical losses due to two-photon absorption (TPA) and free-carrier absorption (FCA) were neglected. The nonlinear coefficient of the $TE_{00}$ mode of the waveguide was set to be 1 $W^{-1}m^{-1}$. Eq. (1) was solved numerically with the split-step Fourier method (*5*).

**OPA experiment details**

The setup of the OPA was designed to operate in either phase-insensitive or phase-sensitive mode. For the PSA mode, phase-correlated signal, pump and idler are needed at the input of the $Si_3N_4$ waveguide. Here, optical processing of the signal and pump was implemented before the OPA to realize this objective. The pump (27 dBm, amplified wave from a tunable C-band laser) was combined with a co-polarized signal via a wavelength-division multiplexer (WDM) and launched into a highly nonlinear fiber (HNLF). An idler was generated at the output via four-wave mixing with a CE of about -15 dB. This copier scheme was used due to its flexibility of signal wavelength tuning and high SNR of the generated idler. The relative phase of the three waves entering the $Si_3N_4$ waveguide dictates the magnitude of the gain in the PSA. The pump was then separated from the signal and idler using another WDM coupler and amplified by a high-power EDFA. A piezoelectric stretcher was used as an optical phase shifter to adjust the phase of the pump. The signal, pump and idler waves were combined by another WDM coupler and coupled into the waveguide with a lensed fiber. We utilized an optical processor (Finisar WaveShaper,) to alternate between PIA and PSA mode, as well as equalizing the power of the signal and idler for PSA. For the PIA with signal (idler) on, the idler (signal) was blocked by the optical processor. The polarizations of the pump, signal and idler were aligned to TE polarization of the waveguide. An optical notch filter was used to attenuate only the high-power pump at the waveguide output, and the output spectra were recorded by an optical spectrum analyzer (OSA, AQ6375B, Yokogawa). Since the pump and signal waves propagated through different optical paths, the relative phase between the pump and signal waves was not constant and slowly varying due to random temperature change and vibrations. A small portion of the amplified signal was filtered out and used as an error signal in a phase locked loop (PLL) with the stretcher compensating the random phase shift to maintain maximum PSA gain.
Since the OPA spectrum is symmetric around the pump wavelength, the signal wavelength was varied only on the short wavelength side of the pump wavelength in the PIA experiment. The OSNR at the input to the waveguide was larger than 46 dB (0.1 nm bandwidth) while it was about 37 dB at the output. This ensured that the extracted NF is accurate and that the input signal can be nearly considered shot-noise limited. In the experiments, we recorded the fiber-to-fiber gain and NF, and we estimated the on-chip gain and NF by accounting for the average coupling loss (2.5 dB/facet) between the lensed fiber and $Si_3N_4$ waveguide. The NF of the parametric amplifier was calculated using the following formula (*6*),

$$NF = \frac{1}{G} + 2\frac{P_{ASE}}{Gh\nu B}, \tag{2}$$

where $G$ is the OPA gain, $P_{ASE}$ is the power of amplified spontaneous emission at the OPA output spectrum with a resolution bandwidth ($B$) equivalent to 0.1 nm, $h$ is the Plank constant and $\nu$ is the frequency of the signal laser. When taking the limited OSNR of input signal wave into account, the $P_{ASE}$ cannot be exactly obtained directly from the OPA output spectrum since the measured noise power at the OPA output spectrum includes both amplified spontaneous emission of the OPA and the amplified noise originally coming from the noise of input signal wave. In our experiment, the noise of input signal wave is mainly from the residual of EDFA amplified spontaneous emission, and the noise is similar at signal and idler wavelength. Therefore, we calculate the $P_{ASE}$ using the following formula,

$$P_{ASE} = P_{noise}^{out} - (G + \eta) \cdot P_{noise}^{in}, \quad (3)$$

where $\eta$ is the OPA conversion efficiency, $P_{noise}^{out}$ is the noise power at the OPA output spectrum with a resolution bandwidth of 0.1 nm and $P_{noise}^{in}$ is the noise power at input port at the signal and idler wavelength with a resolution bandwidth of 0.1 nm. Here, the measured NF at 1554 nm is 1.35 dB if the input signal wave is considered shot-noise limited and is 1.06 dB if the OSNR of input signal wave of 46 dB is taken into account. Considering both uncertainty of OSNR of input signal wave and fluctuation of noise power of OPA at output spectrum, we conclude the on-chip NF is 1.2±0.4 dB.

## Optical transmittance of Si₃N₄ waveguide

Linear and nonlinear losses are key limiting factors for parametric amplifiers since they not only reduce the pump power but also degrade the NF (*7*). In highly nonlinear fibers (HNLFs), this is exemplified by stimulated Brillouin scattering (SBS) which limits the maximum launch pump power. Nonuniform strain can be introduced to reduce the SBS threshold while simultaneously introducing undesired nonuniform dispersion. In highly nonlinear integrated platforms, such as Si, TPA, three-photon absorption and FCA are the nonlinear losses that limit the maximum launch pump power (*8, 9*). However, Si₃N₄ waveguides do not suffer from these effects due to their large bandgap, and their SBS threshold is much higher than SiO₂ (*10*). We investigated the loss in a 1.42m-long Si₃N₄ waveguide by coupling different pump powers to the waveguide and monitoring the output pump power. Fig. S2 shows the on-chip output pump power versus the on-chip input pump power, and the red line indicates 2 dB loss in the chip. Within the tested pump power range, no nonlinear loss was observed, although a 2 dB linear throughput loss was measured.

## Parametric amplification in Si₃N₄ waveguide

Based on the waveguide dispersion measured with OFDR (*11*), we find that the GVD of the waveguide utilized in the experiments of parametric amplification is ~ -36 ps²/km and almost flat in the C-band (1530 – 1560 nm). Consequently, only second-order dispersion is considered, and higher orders of dispersion are neglected in our simulations. For the waveguide used in our PIA experiments, we fit the recorded on-chip gain and conversion efficiency spectra with simulation results shown in Fig. S3. The pump wavelength used in simulations and experiments is 1563 nm, and the waveguide loss is 1.4 dB/m. The estimated experimental pump powers coupled into the chip shown in Fig. S3a-c are 34.4 dBm, 34.4 dBm and 26.8 dBm, while the pump powers used in the simulation are 31.5 dBm, 31.5 dBm and 23.5 dBm, respectively. We found that a GVD of -45 ps²/km leads to a better fit between the simulated and experimental results. No signal gain is obtained with the lowest pump power. Therefore, only the CE spectra are presented in Fig S3c. As a result of the relatively large GVD, the bandwidth of the parametric amplifier with an on-chip gain above 0 dB is about 20-30 nm depending on the pump power. The effective pump power of TE$_{00}$ mode is 2.9 dB smaller than that estimated at the waveguide input according to the experimental and simulation results.

**Noise figure of PSA**

It is known that the waveguide loss impacts the NF of OPAs. An analytical solution has been developed to evaluate the NF of PIA with a lossy waveguide. Approximating the parametric gain coefficient to be constant along the waveguide, we can use the theory of (*12*) to find an expression for the NF of a parametric amplifier in a lossy waveguide as

$$NF_{PIA} = K \frac{(G_{PIA} - 1)}{G_{PIA}} + 1 = \frac{\ln G_{PIA} - 2\ln A}{\ln G_{PIA}} \frac{(G_{PIA} - 1)}{G_{PIA}} + 1. \qquad (1)$$

where $G_{PIA}$ is the PIA gain and $A = \exp(-\alpha L)$ is the linear loss of the waveguide. The factor $K$ ($>1$) was introduced in (*12*), Eqs. 2.14-15 to account for the excess noise from the distributed gain and loss. However, a theoretical expression of the NF of PSA taking into account the linear loss has not been reported earlier. Based on the characteristics of PSA, we here derive a simple equation to analyze the NF of PSA with a linear lossy waveguide. Assuming that the input signal is shot-noise limited, the NF is defined as $NF = SNR_{in}/SNR_{out}$, where $SNR_{in}$ ($SNR_{out}$) denotes the input (output) electrical signal-to-noise ratio of a detected optical signal. For a PIA, the input optical field consists of a signal and a pump. In a PSA, the signal, pump and idler waves are simultaneously entering the amplifier. The power of the signal and idler are assumed to be equal. Here we consider the case in which the phase-sensitive gain of the signal is maximized. A PSA with the same pump power as for a PIA increases the gain of the optical signal by up to 6 dB compared to a PIA while maintaining the output optical noise at the same level as that of the PIA (*13*). As a result, we obtain the following relationship

$$\frac{NF_{PSA}}{NF_{PIA}} = \frac{SNR_{in,PSA}}{SNR_{out,PSA}} \bigg/ \frac{SNR_{in,PIA}}{SNR_{out,PIA}} = 2\frac{|E_{s,out,PIA}|^2}{|E_{s,out,PSA}|^2} = \frac{2G_{PIA}}{G_{PSA}}, \qquad (2)$$

where $G_{PIA}$ and $G_{PSA}$ are the gain of the OPA operated in PIA and PSA modes, respectively. The factor of 2 in Eq. (2) originates from assuming that the input power of the signal and of the idler in the PSA, are both equal to the input signal power of the PIA with the same pump power. Therefore, the input SNR of the signal of the PSA is twice that of the PIA's (*14*). Next, we need to relate the PIA and the PSA gain in the lossy waveguide case. We denote $u = u_0 \exp(-\alpha L/2)$ and $v = v_0 \exp(-\alpha L/2)$ as the total amplitude PIA gain of the signal, and amplitude conversion efficiency of the idler, respectively, where $u_0$ and $v_0$ are the 'lossless' coefficients related by the well-known relation (*15*)

$$|u_0|^2 - |v_0|^2 = 1. \qquad (3)$$

For a PIA, the output signal can be expressed as $E_{s,out,PIA} = uE_{s,in}$ and the gain of the amplifier is

$$G_{PIA} = |u|^2 = |u_0|^2 A. \qquad (4)$$

In the PSA, the idler ($E_{i,in}$) is set to be the conjugate of the signal ($E_{s,in}$) at the input of the PSA, as $E_{i,in} = E_{s,in}^*$. The output signal of PSA is $E_{s,out,PSA} = uE_{s,in} + vE_{i,in}^* = (u + v)E_{s,in}$. Due to the coherent superposition, the PSA gain can be written as (15)

$$G_{PSA} = (|u| + |v|)^2 = (|u_0| + |v_0|)^2 A. \tag{5}$$

Based on the above, the relationship between PIA and PSA gain can be found to be

$$G_{PSA} = (\sqrt{G_{PIA}} + \sqrt{G_{PIA} - A})^2, \tag{6}$$

and by inverting this expression we obtain

$$G_{PIA} = \frac{1}{4}(\sqrt{G_{PSA}} + \frac{A}{\sqrt{G_{PSA}}})^2. \tag{7}$$

To find an expression for the NF of the PSA, we can use Eq. (2), together with the NF expression of the PIA from Eq. (1). After expressing $G_{PIA}$ in terms of $G_{PSA}$ from Eq. (7), we obtain the final expression

$$NF_{PSA} = \frac{1}{2}\left[(1 + K)\left(1 + \frac{A}{G_{PSA}}\right)^2 - \frac{4K}{G_{PSA}}\right], \tag{8}$$

where

$$K = 1 - \frac{2\ln A}{2\ln(G_{PSA} + A) - \ln(4G_{PSA})}. \tag{9}$$

In the lossless case with gain >>1, we note that $NF_{PIA}$ approaches 3 dB and $NF_{PSA}$ approaches 0 dB, as expected. The NF of both PIA and PSA with different waveguide losses can now be calculated. Figure S4a presents the gain difference ($\Delta G = 10\log_{10}(G_{PSA}/G_{PIA})$) between PSA and PIA versus the PIA gain with different waveguide losses. The length of the waveguide is 1.42 m. The blue, red, yellow, purple and green solid lines correspond to the propagation losses of 0, 0.5, 1.5, 2.5 and 3.5 dB/m, respectively. As is shown, $\Delta G$ increases monotonically with the PIA gain. This is due to the coherent addition of the signal and idler as the PIA gain increases. $\Delta G$ saturates at 6 dB at high gain since the power of the idler approaches that of the signal for high gain in the case of the PIA. Figure S4b shows the calculated on-chip NF versus the gain ($G$) of the PIA and the PSA with different propagation losses. Here, we use $G$ as the gain of the waveguide OPA irrespective of whether it is a PIA ($G = G_{PIA}$) or a PSA ($G = G_{PSA}$). The NF of the PIA increases with the gain and approaches the quantum-limit of 3 dB when the waveguide loss is small (i.e., less than 0.5 dB/m) at high gain. This is a typical characteristic of PIAs. For waveguides with higher losses, i.e., >1.5 dB/m, the NF of PIA increases in the small gain regime, reaches a peak, and is reduced at higher gain. When the PIA gain is small and insufficient to compensate the waveguide loss, the NF of PIA is dominated by the linear attenuation of the waveguide and further degraded by spontaneous emission of the PIA (7). When the PIA gain is large enough, the spontaneous emission becomes the main part of the contribution to the NF. It is clear that the NF of the PIA can still be close to 3 dB even for high loss waveguides if the PIA gain is high enough.

For the PSA, the evolution of the NF is quite different. The NF decreases monotonically with the on-chip gain. For a waveguide loss of 3.5 dB/m, the NF of a PSA can be less than 3 dB if the PSA gain is higher than only 5 dB. It can be seen from Fig. S4b that a further increase of the PSA gain to 25 dB can lead to a NF of 1 dB for such a waveguide loss. If the waveguide loss is reduced from 3.5 dB/m to 0.5 dB/m, the decrease in the NF can be reduced from 2.15 dB to 0.35 dB for a PSA gain of 10 dB. Hence the improvement of waveguide loss is of great importance for reducing the NF of waveguide PSAs. In a waveguide with higher $\gamma$, the device length can be reduced for a given gain target, which in turn will have an impact on the NF in a lossy waveguide. As an example, we find that with an increase of $\gamma$ from 1 (Wm)$^{-1}$ to 1.5 (Wm)$^{-1}$, the NF is reduced from 0.6 dB (as shown in Fig. S4b) to 0.33 dB with a gain target of 20 dB and waveguide loss of 1.5 dB/m. In practical applications, the black-box NF consists of the on-chip NF as well as the input coupling loss from the fiber to the waveguide. Considering a low coupling loss of 0.3 dB that can be achievable for CMOS-compatible silicon nitride waveguides (*16*), a ~30 dB-gain PSA with a NF below 1 dB is realistic for waveguide losses less than 2.5 dB/m, which would outperform fiber-based PSAs with record-low NFs (*17*). If the waveguide loss is 1.5 dB/m (similar to our fabricated devices), at 30 dB gain one can expect a NF of well below 0.5 dB.

## Phase-sensitive processing in nonlinear silicon nitride waveguide

To investigate the phase-sensitivity of the $Si_3N_4$ PSA, we use a coherent light source containing the signal, pump and idler. Figure S5 shows a schematic diagram of the experimental setup. The coherent light source (frequency comb with a repetition rate of 25 GHz) is generated by electro-optic modulation of a CW laser (Toptica CTL) (*18*), being spectrally shaped with an optical processor (Finisar Waveshaper) and amplified by a high-power EDFA. The optical processor is used to adjust the power and relative phase of the pump, signal and idler. Since the pump, signal and idler here propagate in the same optical path, they are robust to ambient perturbations such as temperature variations and mechanical vibrations. The signal and idler wavelengths were offset ±4 nm from the pump wavelength, respectively. Other tones of the EO comb were suppressed by the optical processor. The power of the signal and idler were set to be the same and -25 dB lower than that of the pump before the optical waves entered the waveguide. In this way, the extinction ratio between the signal power and noise floor in the output spectrum of the high-power EDFA was more than 20 dB and high enough to observe the nonlinear interference between signal and idler. The relative phase ($\Delta\varphi$) of the input signal was varied from 0 to $2\pi$ by the optical processor with a step of $0.1\pi$. The output optical field of the waveguide was attenuated by 20 dB first and then recorded by an optical spectrum analyzer (OSA, AQ6375B, Yokogawa) after which the relative power of the output signal versus input signal phase was calculated.

## On-chip optical signal processing

We investigated the optical signal processing with the 1.42 m nonlinear $Si_3N_4$ waveguide. Wavelength conversion of a 10 Gbps non-return-to-zero (NRZ) optical signal was performed. Figure S6 presents the schematic diagram of the experiment. We used a commercial Mach-Zehnder modulator (MZM) to generate a 10 Gbps NRZ optical signal. The MZM was biased at the quadrature point. The wavelengths of the signal and pump were set to be 1553 nm and 1563 nm, respectively. An on-chip pump power of about 33.5 dBm was chosen. The pump power was in this experiment somewhat lower than that used in the main text resulting in a wavelength conversion efficiency of about 78%. The power of the 10 Gbps NRZ optical signal was 5.6 dBm before entering the input lensed fiber. The power of the signal and idler in output lensed fiber were 1.1 dBm and -0.5 dBm, respectively. Hence, we obtained an on-chip PIA gain and CE of about 2.5 dB and -1.1 dB, respectively. Benefitting from such a high CE and idler power, the idler can be received directly without any optical pre-amplification. The signal and idler were separated after the $Si_3N_4$ chip by a set of optical filters (ONF, optical notch filter, OBPF, optical band-pass filter) and detected by a commercial photo receiver individually. We used a signal analyzer (BERT N4903A, Agilent) to record the eye diagram of received electrical signal and the bit error rate of the chip-based photonic transmission link. A variable optical attenuator (VOA) was used to adjust the optical power into the receiver for the characterization of the optical transmission link.

Figure S7a-c shows the eye diagram B2B, after-chip and converted signal, respectively. The received optical power was -16 dBm during the measurement of eye diagram. As can be seen, the quality of the signal after the processing of the chip is slightly degraded. Regarding the WC, the idler is more degraded than the signal. In addition, we measured how the BER changed with the ROP of different optical signals. The BER curve is depicted by Fig. S7d. For a BER of $10^{-6}$, the penalties of the signal and idler propagating through the chip were around 0.3 dB and 1 dB, respectively. The instability of coupling under high power and minor spectral ripples may result in the degradation of signal and idler after transmission through the chip. The instability of coupling may contribute more to the degradation of idler than to signal, since the idler is related to the product of signal and pump so that it is more sensitive than the signal during the wavelength conversion (*19*).

**Fig. S1**

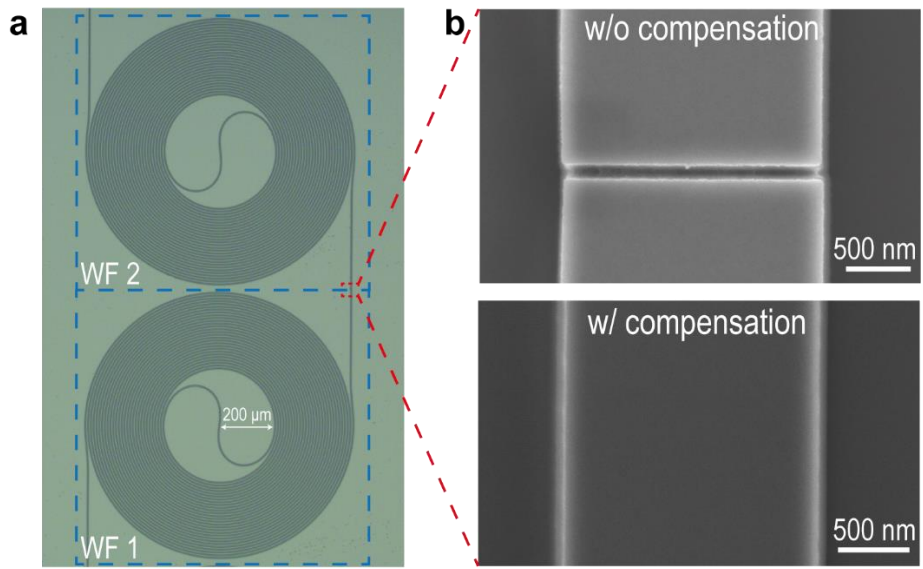

**Fig. S1. Images of spiral waveguides. a**, Optical microscope image of two spiral waveguide units. The blue dashed boxes depict two adjacent writing fields (WFs), and the red dashed box indicates the location of the stitching error. **b**, Scanning electron microscopy images of the resist exposed by EBL with and without stitching error compensation.

**Fig. S2.**

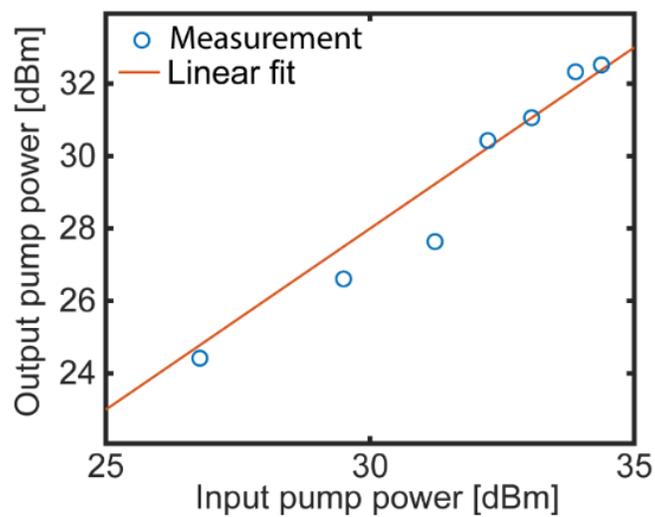

**Fig. S2.** The on-chip output pump power versus the on-chip input pump power. The red line indicates 2 dB linear propagation loss.

**Fig. S3.**

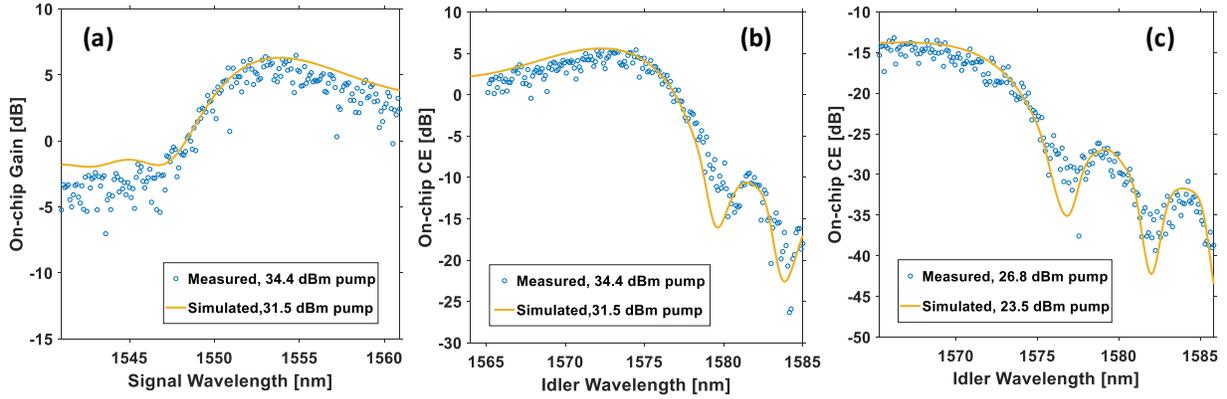

**Fig. S3.** On-chip (a) gain and (b) CE spectra with a 34.4 dBm pump power. (c) On-chip CE spectra with a 26.8 dBm pump power in the experiment.

**Fig. S4.**

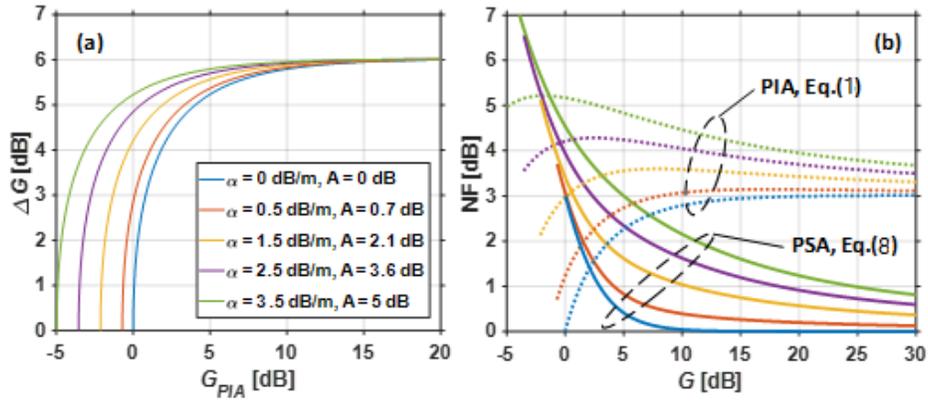

**Fig. S4.** (a) The gain difference between PSA and PIA varying with PIA gain. (b) Chip NF versus parametric gain for both PIA and PSA. The waveguide length is 1.42m.

**Fig. S5.**

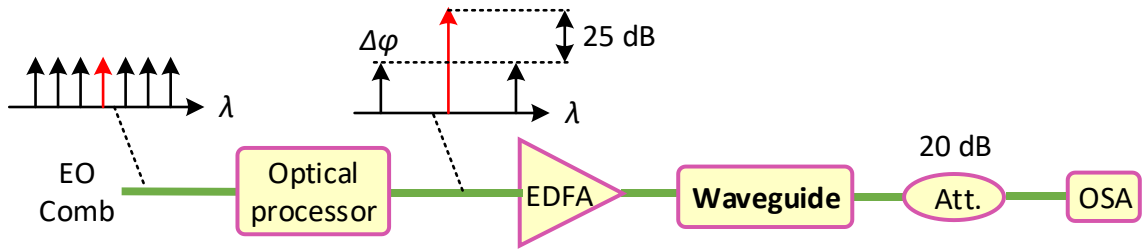

**Fig. S5.** Schematic diagram of the experimental setup to study the phase-sensitive process in a nonlinear silicon nitride waveguide.

**Fig. S6.**

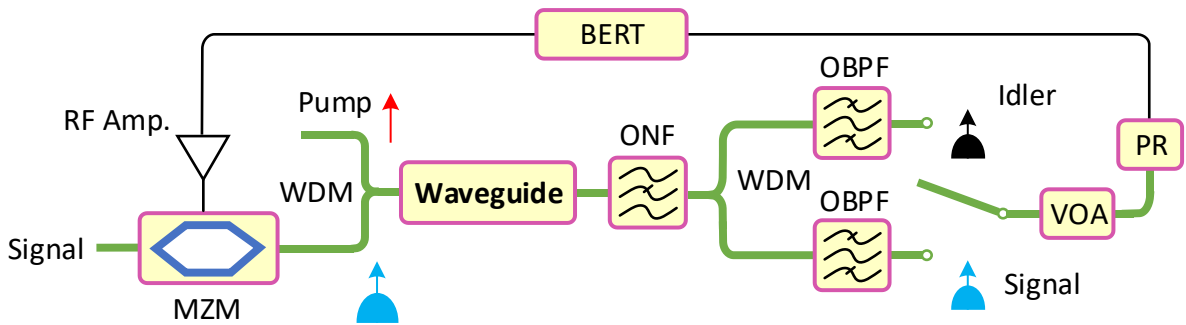

**Fig. S6.** Schematic diagram of wavelength conversion of 10 Gbps NRZ signal.

**Fig. S7.**

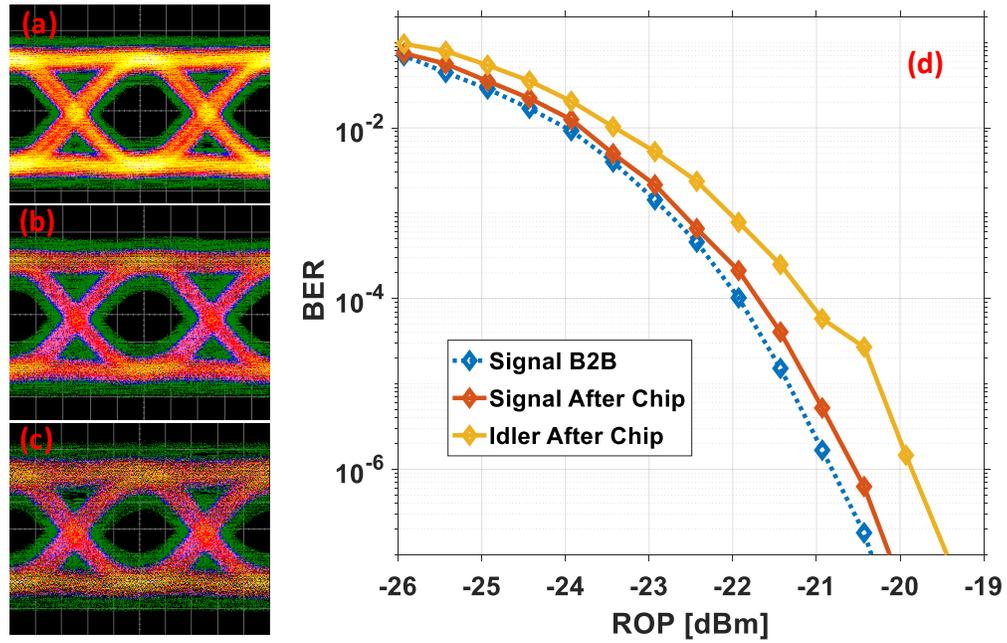

**Fig. S7.** Eye diagrams of (a) B2B, (b) after-chip and (c) converted signals with a received optical power (ROP) of -16 dBm. (d) BER changing with ROP.